\documentstyle[12pt,epsfig]{article}
\def\e{\epsilon}

\def\ycut{y_{\rm cut}}
\def\ymin{y_{\rm min}}

\def\smin{s_{\rm min}}
\begin{document}
\pagestyle{plain}

\begin{titlepage}
\vspace*{-1cm}
\begin{flushright}
DTP/97/24   \\
DESY-97-084 \\
May 1997 \\
Revised August 1997\\
\end{flushright}                                
\vskip 1.cm
\begin{center}                                                             
{\Large\bf
Radiative Corrections to the \\[2mm]
Photon +~1 Jet Rate at LEP}
\vskip 1.3cm
{\large A.~Gehrmann--De Ridder$^a$, T.~Gehrmann$^b$ and  E.~W.~N.~Glover$^a$}
\vskip .2cm
$^a$~{\it Department of Physics, University of Durham,
Durham DH1 3LE, England }\\
$^b$~{\it DESY, Theory Group, D-22603 Hamburg, Germany}
\vskip 2.3cm   
\end{center}      
\begin{abstract}
We present the results of a calculation of the rate for the annihilation
 of an $e^+e^-$ pair into a jet containing an energetic photon 
and a single further jet at order $\alpha \alpha_{s}$.
By comparing these results with the existing data 
on the photon +~1 jet rate from the ALEPH Collaboration, 
we make a first determination of the process independent quark-to-photon 
fragmentation function 
$D_{q \to \gamma}(z,\mu_{F})$ at order $\alpha \alpha_{s}$.
We also compare our prediction for the inclusive photon energy spectrum
with the recent OPAL measurement.
\end{abstract}                                                                
\vfill
\end{titlepage}
\newpage                                                                     

The production of final state photons at large transverse momenta is 
one of the key observables studied in hadronic collisions. 
Data on high-$p_T$ photon production  
yield valuable information on the gluon distribution in the proton. 
Moreover, if the Standard Model Higgs boson has a mass lying
below the $WW$-threshold, one of the best 
detection channels at the CERN LHC is 
via its two photons decay mode~\cite{LHC}. 
A precise understanding of the background processes yielding 
two photon final states is therefore crucial 
for Higgs searches at the LHC. 

Photons produced in hadronic collisions arise essentially 
from two different sources: `direct' or `prompt' photon production
via the hard partonic processes such as
$qg \to q \gamma$ and $q\bar q \to g \gamma$ 
or through the `fragmentation' of a hadronic jet into a single photon 
carrying a large fraction of the jet energy.
The former
gives rise to perturbatively calculable short-distance contributions 
whereas the latter is primarily a long 
distance process which cannot be calculated perturbatively and
is described in terms of the quark-to-photon 
fragmentation function~\cite{phofrag}. 
In principle,
this fragmentation contribution could be suppressed to a certain
extent by imposing isolation cuts on the photon, typically formulated 
by limiting the amount of hadronic energy allowed within an isolation 
cone around the photon. 
However, the matching of experimental isolation cones onto a 
theoretical calculation beyond the lowest order level is far from 
trivial~\cite{cone,giele}, and it has been shown that a sizeable 
contribution to the isolated 
photon cross section
from the quark-to-photon fragmentation process 
remains even after isolation cuts have been applied~\cite{Vogelsang}. 

All estimates of the fragmentation
contribution to the production 
of one or two final state photons in hadron collisions
so far~\cite{allgamma} 
have had to rely on a model of the photon fragmentation function.
This function 
obeys a perturbative evolution equation with a non-perturbative 
boundary condition usually parameterized in the form of an initial 
distribution at some low starting scale $\mu_0$. Several models,
based either on an asymptotic solution of the evolution 
equation~\cite{owens} or on estimates of the non-perturbative 
boundary conditions inspired by Vector Meson Dominance models~\cite{grv}
can be found in the literature. 
It is clear that the lack of a
precise knowledge on the quark-to-photon fragmentation function introduces a 
{\it non-quantifiable} systematic uncertainty into any 
theoretical calculation of photon production cross sections. 
A precise determination of the process independent
quark-to-photon fragmentation 
function is therefore needed and 
the results obtained in the remainder of this letter should help to
achieve this.

A first measurement of the quark-to-photon fragmentation function 
has been performed~\cite{aleph} by the ALEPH collaboration at CERN
from an analysis of two jet events in which one of the jets contains a 
photon carrying a large fraction ($z>0.7$) of the jet energy. 
In these `photon' +~1 jet events, 
a democratic approach is used to group the particles into jets
and the photon is clustered simultaneously with the other hadrons.
A comparison between this measured rate 
and the calculated rate up to ${\cal O}(\alpha)$ 
\cite{andrew} yielded a first  
determination \cite{aleph} of the quark-to-photon fragmentation function 
accurate at this order.
Both theoretical and experimental analyses 
used the Durham jet algorithm \cite{durham}.
Note that $z$, the fraction of 
electromagnetic energy within the `photon' jet, 
is in general different from the parameter $x_{\gamma}$ considered 
in inclusive photon production~\cite{KT}. In the collinear core
of the `photon jet' however, where the non-perturbative effects 
arise, the definitions of these two parameters coincide.

In the present letter, we report the results of a calculation of the
${\cal O}(\alpha \alpha_s)$ QCD corrections to the 
`photon' +~1 jet rate, performed using the same democratic approach.
Together with the published data \cite{aleph}, 
this enables us to obtain a first determination 
of the process independent quark-to-photon fragmentation function 
at ${\cal O}(\alpha \alpha_{s})$. 
The details of this calculation will be presented in a subsequent  
publication~\cite{big}.

A common feature of all model estimates of the 
quark-to-photon fragmentation function~\cite{owens,grv} is the resummation 
of leading $(\alpha_s^n \log^{n+1} \mu_F^2)$ and subleading  
$(\alpha_s^n \log^n \mu_F^2)$
logarithms of the mass factorization scale $\mu_F$ to all orders in 
the strong coupling constant. The results of this resummation
procedure
seem to suggest that the 
photon fragmentation function 
is ${\cal O}(\alpha/\alpha_s)$ and that quark and gluon fragmentation into 
photons contribute at the same order. 
This view is widely held for the photon 
fragmentation as well as for the 
photon structure function. 
However, in the region of high $z$ and where jets are resolved, this is 
not necessarily the correct procedure to adopt.
For example, the distribution of electromagnetic energy within the photon jet of photon + 1 jet events, for a single quark of charge $e_q$, can be written \cite{andrew},
\begin{equation}
\frac{1}{\sigma_0} \frac{d\sigma}{dz}
= 2 D_{q\to \gamma}(z,\mu_F) + \frac{\alpha e_q^2}{\pi} 
P_{q \gamma}^{(0)}(z) \log \left(\frac{s_{\rm cut}}{\mu_F^2}\right) + \ldots,
\label{eq:sig0}
\end{equation}
where $\ldots$ represents terms where the photon is perturbatively
isolated  (proportional to $\delta(1-z))$ and other 
perturbative contributions which are
well behaved as $z \to 1$.
The non-perturbative fragmentation function is a solution of,
\begin{equation}
\frac{\partial D_{q\to \gamma}}{\partial \log \mu_F^2} = 
\frac{\alpha e_q^2}{2 \pi} P_{q \gamma}^{(0)}(z) + \ldots ,
\end{equation}
and $s_{\rm cut}$ is determined by the jet clustering algorithm.  
In the Durham jet clustering algorithm and at large $z$, $s_{\rm cut} 
\sim s z(1-z)^2/(1+z) \sim p_T^2$ \cite{andrew} where $p_T$ is the 
transverse momentum of the photon with respect to the cluster.
In the usual approach, the solution of the evolution equation for 
$D_{q\to \gamma}$, is proportional to $\log(\mu_F^2/\mu_0^2)$ where 
$\mu_0$ is some small scale and the conventional assignment of a power 
of $1/\alpha_s$ to the fragmentation function is clearly motivated.
For `physical' scale choices, $\mu_F^2 \sim s_{\rm cut}$,\footnote{For 
inclusive processes, $s_{\rm cut} \sim  s$ is more appropriate.} and the 
perturbative contribution  is suppressed relative to the fragmentation 
contribution, so that for $z < 1$, the fragmentation function is indeed 
more significant.
However, at large $z$, we see that  the transverse size of the photon jet 
cluster decreases such that $s_{\rm cut} \to 0$.
The hierarchy $s_{\rm cut} \sim \mu_F^2$ and $\mu_F^2 \gg \mu_0^2$ is
no longer preserved and both contributions in
eq.~(\ref{eq:sig0}) are important.
Large logarithms of $(1-z)$ are the most dominant contributions.
Since, from the experimental point of view,
we are required to observe isolated or almost isolated photons, 
this is precisely the region that we are most interested in.
For this reason, we choose not to impose the conventional prejudice and 
resum the logarithms of $\mu_F$ {\em a priori}.   
Instead, we work within a fixed order framework, to isolate the relevant 
large logarithms.

As mentioned above,   
potentially large logarithms of $(1-z)$ are present and need to be resummed.
At present however, we have not accomplished this resummation and,
to avoid unstable behaviour at large $z$, we choose a form of the 
fragmentation function that renders the calculation well behaved as $z \to 1$.
This approach is identical to the procedure applied in the 
lowest order calculation~\cite{andrew} and in the lowest order
experimental analysis~\cite{aleph}.
It is thus possible to 
make a direct comparison of our next-to-leading order results with 
these earlier leading order results, and to test 
the stability of the perturbative 
fixed order approach.

\begin{figure}[t]
\begin{center}
~ \epsfig{file=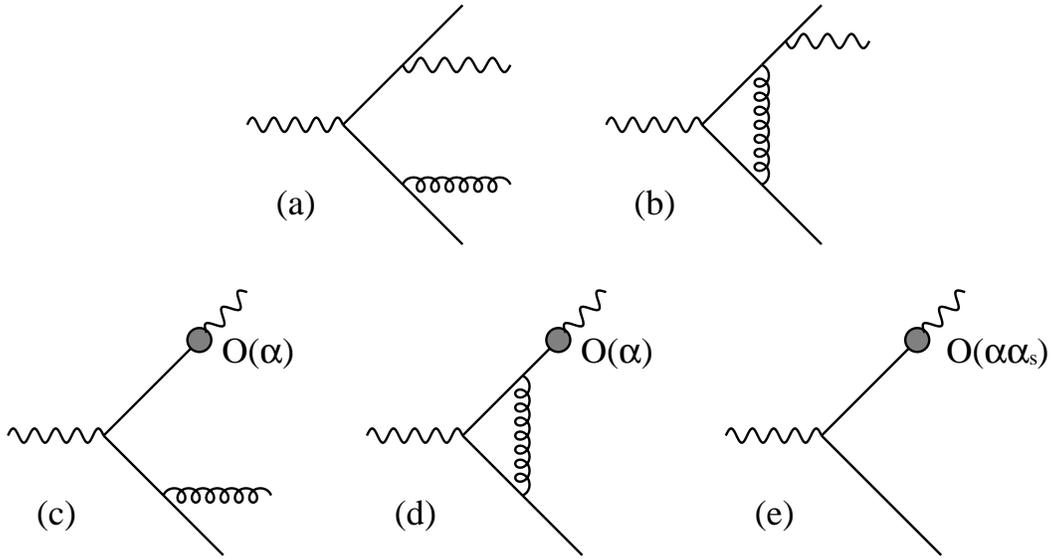,width=14cm}
\caption{Parton level subprocesses contributing to the photon +~1 jet 
rate at ${\cal O}(\alpha\alpha_s)$.} 
\label{fig:class}
\end{center}
\end{figure}
The photon +~1 jet rate in $e^+e^-$ annihilation at ${\cal O} (\alpha 
\alpha_s)$
receives contributions from five parton-level subprocesses
displayed in Fig.~\ref{fig:class}:
\begin{itemize}
\item[{(a)}] The tree level process $\gamma^* \to q \bar{q}g\gamma$,
where the final state particles are clustered together such that a
``photon jet'' and one additional jet are observed in the final state.
\item[{(b)}] The one loop gluon correction to the $\gamma^* \to q
\bar{q}\gamma$ process, where the photon and one of the quarks are clustered
together or the photon is isolated while both quarks form a single jet.
\item[{(c)}] The process $\gamma^* \to q
\bar{q}g$, where one of the quarks fragments into a photon while
the remaining partons form only a single jet.
\item[{(d)}] The one loop gluon correction to the $\gamma^* \to q
\bar{q}$ process, where one of the quarks fragments into a photon.
\item[{(e)}] The tree level process $\gamma^* \to q\bar{q}$ with a
generic ${\cal O}(\alpha \alpha_{s})$ counterterm present in the {\it bare}
quark-to-photon fragmentation function.
\end{itemize}

Although the `photon' +~1 jet cross section 
is finite at ${\cal O}(\alpha \alpha_{s})$, 
all these contributions  
contain divergences (when the photon 
and/or the gluon are collinear with one of the quarks, 
when the gluon is soft or because 
the bare quark-to-photon fragmentation 
function contains infinite counter terms).
These divergences need to be isolated and cancelled analytically.
It is only after this cancellation has taken place, that it is possible 
to numerically evaluate
the remaining finite contributions in a Monte Carlo program 
containing the experimental jet reconstruction algorithm, and ultimately to  
evaluate the `photon' +~1 jet rate at ${\cal O}(\alpha \alpha_{s})$.

The assignment of a single power of $\alpha$ to the quark-to-photon 
fragmentation function (and ${\cal O}(\alpha\alpha_s)$ to the 
gluon-to-photon fragmentation function)
carries through when factorizing the explicit 
singularities present in the real and virtual processes with photon emission
into the fragmentation functions.
Here,  
the leading perturbative counterterm in the quark-to-photon 
fragmentation function is ${\cal O}(\alpha)$ while the leading counterterm
in the gluon-to-photon fragmentation function is 
${\cal O}(\alpha\alpha_s)$. 
Consequently, the gluon-to-photon fragmentation function plays 
no role at the fixed order ${\cal O}(\alpha\alpha_s)$; 
since radiating the gluon contributes an additional factor of 
$\alpha_s$, it first contributes at ${\cal O}(\alpha\alpha_s^2)$.
In fact, all of the model estimates~\cite{owens,grv} indicate 
that the gluon-to-photon fragmentation is orders of magnitude smaller than the 
quark-to-photon fragmentation function at large $z$ and
 we ignore it throughout.

Following the philosophy of \cite{gg},
we use the concept of a theoretical resolution criterion $\smin$
to define the singular regions.  
Within the singular regions, the matrix elements are approximated
and the unresolved variables analytically integrated.
At next-to-leading order, most QCD processes receive singular 
contributions when either one gluon is soft or two particles are 
collinear -- in other words 
a single particle is theoretically unresolved.
However, the evaluation of the singular contributions 
associated with the process $\gamma ^* \to q \bar{q}g \gamma$ is of 
particular interest as it 
contains various ingredients which could directly be applied 
to the calculation of jet observables at next-to-next-to-leading order.
For example, there are contributions where {\em two} of the 
final state partons are theoretically unresolved.
The three different double unresolved contributions which 
occur in this calculation are: 
the {\it triple collinear} contributions, 
arising when the photon and the gluon are simultaneously collinear 
to one of the quarks, the {\it soft/collinear} contributions 
arising when the photon is collinear to one of the quarks 
while the gluon is soft and the {\it double single collinear} contributions, 
resulting when the photon is collinear to one of the quarks while the gluon 
is collinear to the other.
In addition, there are also single unresolved contributions from  
the one-loop $\gamma^* \to q\bar q \gamma$ process  
and from the tree-level fragmentation process
$\gamma^* \to q\bar q g$ when two of the final state particles are collinear.
A detailed derivation of each of these singular contributions,
along with how the different single and double 
unresolved regions match onto each other,
will be presented in \cite{big}.

The combination of the unresolved contributions present in the 
processes shown in Fig.~\ref{fig:class}(a)-(d)  
yields a result 
that still contains single and double poles in $\e$.
These pole terms however are proportional 
to the universal next-to-leading order splitting function 
$P_{q\gamma}^{(1)}$ ~\cite{curci} and a convolution of two 
lowest order splitting functions, 
$(P_{qq}^{(0)}\otimes P_{q \gamma}^{(0)})$. 
Hence, they are factorized into the next-to-leading order 
counterterm of the bare quark-to photon fragmentation function 
\cite{fac}
present in the contribution depicted in Fig.~\ref{fig:class}(e),
yielding a finite and factorization scale ($\mu_{F}$) dependent 
result \cite{big}.

Once the singularities have been removed, the remaining 
finite contributions can be numerically evaluated. 
Often this is realized within the {\it phase space slicing} method,
first introduced 
by \cite{kramer} and further developed in \cite {gg} for general 
next-to-leading order QCD processes.
The original formulation of the phase space slicing method requires 
the analytical calculation of an approximated matrix element inside the 
theoretically unresolved regions, while the full matrix element 
is evaluated numerically only outside any singular region.
This procedure turns out to be inappropriate   
where more than one particle is potentially unresolved.
Apart from generically double unresolved regions, 
we find areas in the four parton phase space which belong simultaneously 
to two different single unresolved regions.
Those areas are not treated correctly within the phase space slicing procedure.
A consistent treatment is possible in the {\it hybrid subtraction} method 
developed in \cite{eec}. 
This scheme attempts to combine the advantages of the phase space slicing 
procedure with the alternative {\it subtraction} methods \cite{subtract}.
The parton resolution parameter is used to isolate the poles,
but, rather than assuming that the approximated matrix elements 
are exact,  
the difference between the full matrix 
element and its approximation is
numerically evaluated in all unresolved regions.   

The numerical program finally evaluating the `photon' +~1 jet rate 
at ${\cal O}(\alpha\alpha_s)$ contains four separate contributions.
These contributions are 
determined according to the number of partons in the final state and 
the presence or absence of the quark-to-photon fragmentation function.
For each final state,  `photon' +~1 jet events are selected 
using a democratic jet clustering algorithm  
and a given value of the jet resolution parameter $y_{{\rm cut}}$.
The ``fragmentation function'' used in the numerical evaluation of the 
cross section is the sum 
of the $\mu_{F}$-dependent quark-to-photon fragmentation 
function and a finite 
contribution from the unresolved photon region.
The four contributions are:
\begin{itemize}
\item[{(A)}] {\bf 2 partons + photon} \newline
This contribution corresponds to the process $\gamma^*
\to q \bar q \gamma$ with a hard photon in the final state. 
It is present at ${\cal O}(\alpha)$ and ${\cal O}(\alpha\alpha_s)$  
(due to the presence of a theoretically 
unresolved real or virtual gluon). 
\item[{(B)}] {\bf 2 partons + ``fragmentation''} \newline
The   
$\gamma^* \to q \bar q \otimes D_{q\to \gamma}$ process is present 
at ${\cal O}(\alpha)$ and ${\cal O}(\alpha\alpha_s)$.
In particular, it contains the finite terms corresponding to the
$q-\gamma$ collinear region at ${\cal O}(\alpha)$ and the 
finite parts associated  
with all double unresolved regions at ${\cal O}(\alpha\alpha_s)$.
\item[{(C)}] {\bf 3 partons + photon} \newline
This contribution is only present at ${\cal O}(\alpha \alpha_s)$ and 
describes the $\gamma^* \to q \bar q \gamma g$ process where 
both photon and gluon are theoretically resolved.
\item[{(D)}] {\bf 3 partons + ``fragmentation''} \newline
The  
$\gamma^* \to q \bar q g \otimes D_{q\to \gamma}$ process
with a hard gluon in the final state, 
is only present at ${\cal O}(\alpha \alpha_s)$. 
\end{itemize}

\begin{figure}[t]
\begin{center}
~ \epsfig{file=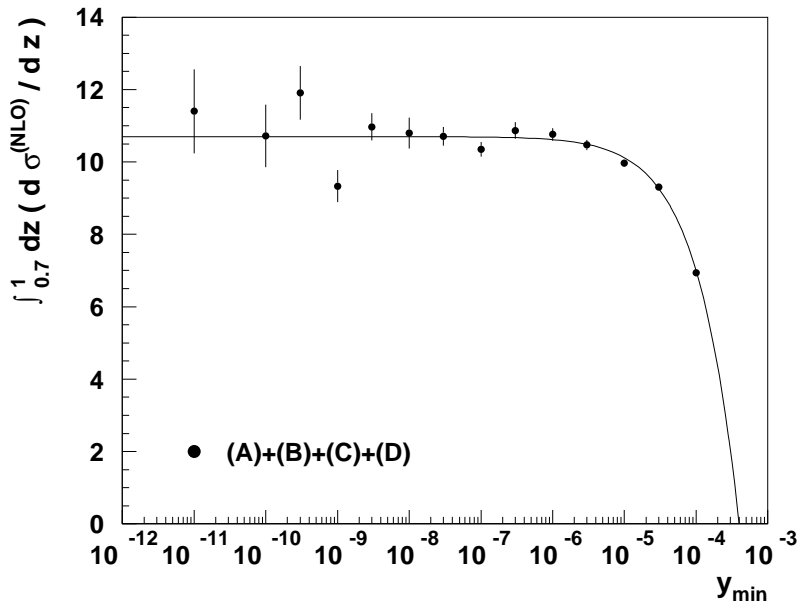,width=8cm,angle=-90}
\caption{Sum of all ${\cal O}(\alpha\alpha_s)$ contributions to the 
total photon + 1 jet rate for a single quark of charge $e_q$ such that
$\alpha e_q^2 = 2\pi$, $\alpha_s (N^2-1)/2N = 2\pi$ using
the Durham  jet algorithm with $y_{{\rm cut}}=0.1$,  
and integrated for $z>0.7$.}
\label{fig:ymin2}
\end{center}
\end{figure}

Each of the contributions listed above depends logarithmically (in fact 
as $\log^3 (y_{{\rm min}})$) on 
the theoretical resolution parameter $y_{{\rm min}}=s_{{\rm min}}/Q^2$, 
which is used throughout the calculation to divide the
final state phase space into different resolved and unresolved 
regions.
However, the physical `photon' + 1~jet cross section,
which is the sum of all 
theoretically resolved and unresolved contributions,
{\it must} of course 
be independent of the choice of $y_{{\rm min}}$.
That this is indeed the case is illustrated in Fig.~\ref{fig:ymin2}
where we show the sum of all ${\cal O}(\alpha\alpha_s)$ contributions
to the total rate as a function of $\ymin$.
For simplicity, we show only the ${\cal O}(\alpha\alpha_s)$
contribution 
for a single quark of charge $e_q$ such that
$\alpha e_q^2 = 2\pi$, $\alpha_s (N^2-1)/2N = 2\pi$ using
the Durham jet algorithm with $y_{{\rm cut}}=0.1$,  
and integrated for $z>0.7$.
We see that the cross section approaches (within 
numerical errors) 
a constant value provided that $y_{{\rm min}}$ is chosen to be 
small enough, indicating 
a complete cancellation of all powers of $\log (y_{{\rm min}})$. This 
provides a strong check on the correctness of our results 
and on the consistency of our approach. 
The behavior at large values of 
$y_{{\rm min}}$ can be understood to be due to terms proportional to
$y_{{\rm min}} \log^2(y_{{\rm min}})$, which have been neglected 
in the analytic evaluation of all
theoretically unresolved contributions. It can clearly be seen that these 
terms are already unimportant at  $y_{{\rm min}}=
10^{-6}$, which will be used in all numerical studies below. 

We can now  determine 
the  
quark-to-photon fragmentation 
function $D(z,\mu_F)$ up to this order 
from a comparison between the measured `photon' +~1 jet rate  \cite{aleph}
and the results of our calculation.
This function, which parameterizes the 
perturbatively incalculable long-distance contribution to the photon
radiation off a quark has to satisfy a perturbative evolution equation 
in the factorization scale $\mu_F$. All unknown long-distance contributions 
can therefore be attributed to the behavior of $D(z,\mu_0)$, 
the initial value of this fragmentation function, which 
is to be fitted to the experimental data, at some 
low scale $\mu_0$. Indeed, the next-to-leading order fragmentation
function can be expressed as an {\it exact} solution of the evolution equation 
up to ${\cal O}(\alpha \alpha_s)$ \cite{big},
\begin{eqnarray}
D(z,\mu_{F})&=&
 D(z,\mu_{0}) + \frac{\alpha e_{q}^2}{2\pi}P^{(0)}_{q \gamma}(z)
\log\left(\frac{\mu^2_{F}}{\mu_{0}^2}\right) 
+\frac{\alpha e_{q}^2}{2\pi} \frac{\alpha_{s}}{2 \pi}
\left(\frac{N^2 -1}{2N}\right)P_{q \gamma}^{(1)}(z)
\log \left(\frac{\mu^2_{F}}{\mu_{0}^2}\right)
\nonumber\\
& & +
\frac{\alpha_{s}}{2 \pi}
\left(\frac{N^2 -1}{2N}\right) \;
\log \left(\frac{\mu^2_{F}}{\mu_{0}^2}\right) P_{qq}^{(0)}(z)\otimes 
\left[\frac{\alpha e_{q}^2}{2 \pi}\frac{1}{2}P_{q \gamma}^{(0)}(z)
\log \left(\frac{\mu^2_{F}}{\mu_{0}^2}\right)
+D(z,\mu_{0})\right].
\nonumber\\
\label{eq:Dnlo}
\end{eqnarray}
It should be re-emphasized that this solution does not
take the commonly implemented (e.g.~\cite{owens,grv}) 
resummations of $\log \mu^2_{F}$ 
into account. As explained above, this appears the only consistent  
procedure since powers of
$\log (1-z)$ may be of equal importance in the large $z$ region~\cite{webber},
which is the main region of interest in the present study. Finally,  
this procedure is identical to the procedure applied in earlier 
theoretical~\cite{andrew} and experimental~\cite{aleph} analyses at 
lowest order, such that a direct and self-consistent 
quantification of the next-to-leading order effects 
in the `photon' +~1 jet rate becomes feasible. 

Furthermore, since this solution is  exact  
at the order under consideration, 
the factorization scale dependence of the `photon' +~1 jet rate is 
eliminated, i.e.~our results are independent of the choice of $\mu_F$. 
\begin{figure}[t]
\begin{center}
~ \epsfig{file=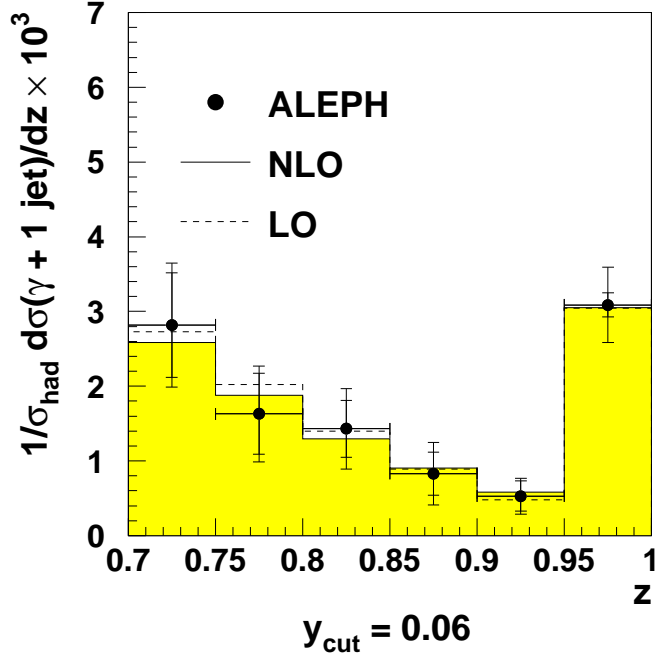,width=9cm,angle=-90}
\caption{Comparison of the `photon' +~1 jet rate at ${\cal O}(\alpha)$ and
${\cal O}(\alpha\alpha_s)$ 
with the ALEPH data ($y_{{\rm cut}}=0.06$), including the fitted 
quark-to-photon fragmentation function.}
\label{fig:y006}
\end{center}
\end{figure}
\begin{figure}[t]
\begin{center}
~ \epsfig{file=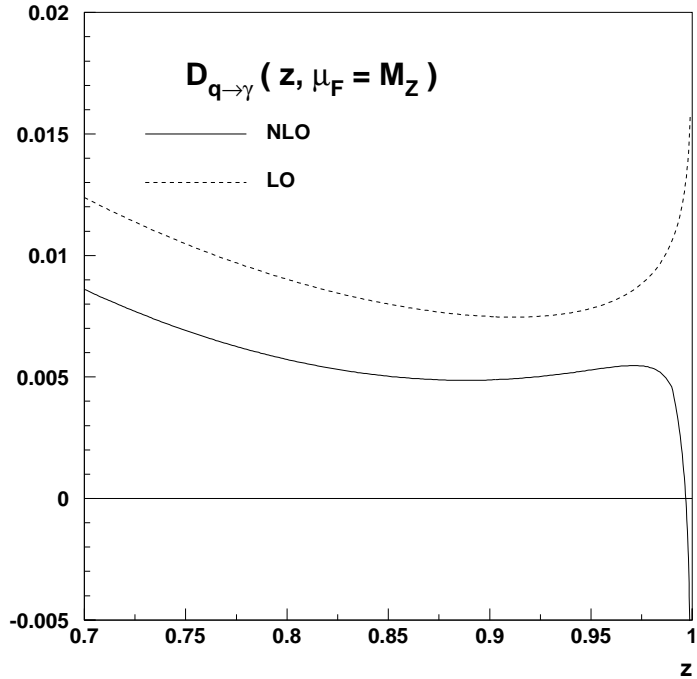,width=9cm,angle=-90}
\caption{The measured quark-to-photon fragmentation function at 
leading
\protect{\cite{aleph}} and next-to-leading (this study) 
order for a quark of 
unit charge. The factorization scale is taken to be $\mu_F=M_Z$.}
\label{fig:frag}
\end{center}
\end{figure}

A three parameter fit 
to the ALEPH `photon' +~1 jet
data \cite{aleph} for the $z$ distribution, 
$\frac{1}{\sigma_{0}}\frac{d\sigma}{dz}$,  
for the jet resolution parameter $y_{{\rm cut}}=0.06$ 
and $\alpha_s(M_z^2) = 0.124$\footnote{At this order, 
all choices of the strong coupling constant are equally valid.  
This value of $\alpha_s$ is chosen so that the 
observed hadronic cross section is reproduced by the ${\cal O}(\alpha_s)$
calculation.}
yields, 
\begin{equation}
D^{NLO}(z,\mu_{0})=\frac{\alpha e_{q}^2}{2 \pi} 
\left(-P^{(0)}_{q \gamma}(z)\log(1-z)^2 \;+\,20.8\,(1-z)-11.07\right),
\label{eq:fitnlo}
\end{equation}
where $\mu_{0}=0.64$~GeV.
For reference, the lowest order fit obtained 
by the ALEPH Collaboration \cite{aleph} is,
\begin{equation}
D^{LO}(z,\mu_{0})=\frac{\alpha e_{q}^2}{2 \pi} 
\left(-P^{(0)}_{q \gamma}(z)\log(1-z)^2 - 13.26\right),
\label{eq:fitlo}
\end{equation}
with $\mu_{0}=0.14$~GeV.
In both cases, the logarithmic term proportional to $P^{(0)}_{q \gamma}(z)$
is introduced to ensure that the predicted $z$ distribution is 
well behaved as $z \to 1$ \cite{andrew}. It should be kept in mind 
that both parameterizations are only fitted to data with $z>0.7$, which 
is thus the natural limit of their validity. 

The ALEPH data and the results of the ${\cal O}(\alpha \alpha_s)$ 
calculation using the fitted next-to-leading order 
fragmentation function are shown in 
Fig.~\ref{fig:y006}. (For reference, we have also displayed the 
${\cal O}(\alpha)$ 
prediction~\cite{andrew} using  
the ALEPH fit of the lowest order fragmentation
function~\cite{aleph}.) 
The fit also provides a good description of the $z$ distribution for other 
values of $\ycut$ \cite{big}.
This is to be expected since the non-perturbative effects are largely in
the collinear core of the `photon' jet which is fully contained within the
jet definition. Moreover, it  turns out that the next-to-leading order     
corrections are moderate for all values of $\ycut$, which clearly
demonstrates the perturbative stability of our fixed order approach.

The resulting next-to-leading order ($\overline{{\rm MS}}$)
quark-to-photon fragmentation function (for a quark of unit charge) 
at a factorization scale $\mu_F=M_Z$ is shown 
in Fig.~\ref{fig:frag} and 
compared with the lowest order fragmentation function obtained 
in~\cite{aleph}. A large difference between the leading and next-to-leading 
order quark-to-photon fragmentation functions is observed only for  
$z$ close to 1, indicating the presence of large $\log (1-z)$. 
\begin{figure}[t]
\begin{center}
~ \epsfig{file=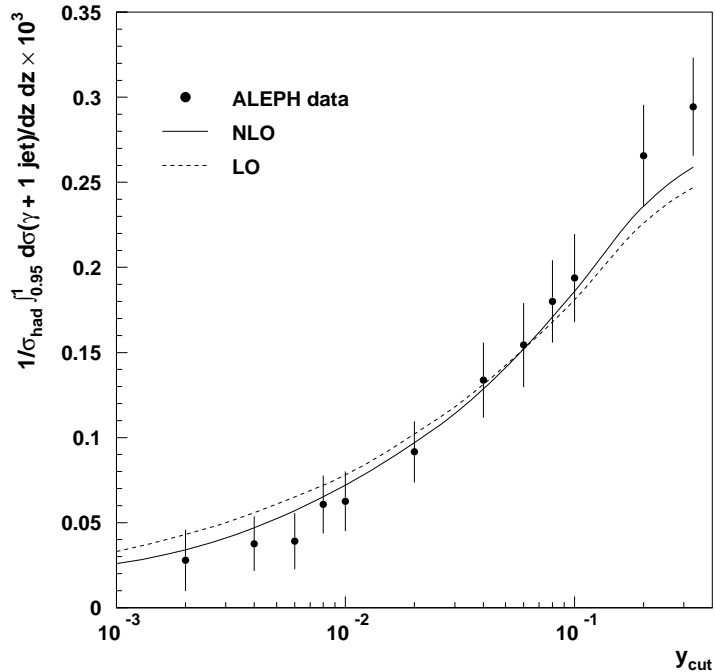,width=9cm,angle=-90}
\caption{The integrated photon +~1 jet rate above $z=0.95$ as function of
$y_{{\rm cut}}$, compared with the ${\cal O}(\alpha)$ 
and ${\cal O}(\alpha \alpha_s)$
order calculations including the appropriate 
quark-to-photon fragmentation functions.}
\label{fig:ycut}
\end{center}
\end{figure}

Having described the $z$ distribution in the measured range, we shall now
focus
only on the `photon' +~1 jet rate for $z>0.95$ 
which can be viewed as the {\it isolated} photon +~1 jet rate 
in the democratic clustering approach \cite{aleph}. 
This division between isolated and non-isolated is motivated both by the
measured $z$ distribution (see Fig.~\ref{fig:y006}) and the observation that
hadronization effects cause the photon jet to have a $z$ value slightly
 less than unity \cite{aleph}. 
In previous theoretical \cite{KT,gs,KS} and experimental \cite{iso} analyses 
of isolated photon +~$n$ jet rates,
the candidate photon was isolated from the other hadrons 
using a geometrical cone, {\it before} these hadrons were clustered into 
jets by the jet recombination algorithm.
The photon was said to be `isolated' if it was accompanied only by
a small (but non-zero) 
amount of hadronic energy inside the cone.  
As a result of these analyses it was found that large negative next-to-leading
order corrections were needed to obtain a reasonable agreement between 
theory and experiment~\cite{gs,gm}.
Interestingly, a lowest order calculation of the photon +~1 jet rate 
for the cone algorithm using the process independent photon fragmentation
function measured by ALEPH \cite{aleph}
and a suitable choice of $z$ to define the isolation yields
good agreement with the data \cite{cone}.

Using the results of our calculation and 
the fitted quark-to-photon fragmentation function, we have 
determined the isolated photon +~1 jet rate 
in the democratic approach up to ${\cal O}(\alpha \alpha_s)$. 
The result of this calculation  compared with 
data from ALEPH~\cite{aleph} and the leading order calculation~\cite{andrew}
is shown in Fig.~\ref{fig:ycut}. It can clearly be seen that inclusion of 
the next-to-leading order corrections improves the agreement between 
data and theory over the whole range of $y_{{\rm cut}}$. 
It is also apparent that the next-to-leading order corrections 
to the isolated photon +~1 jet rate obtained in this 
democratic clustering approach are of reasonable size 
indicating a better perturbative stability of this {\it isolated} 
photon definition. 
\begin{figure}[t]
\begin{center}
~ \epsfig{file=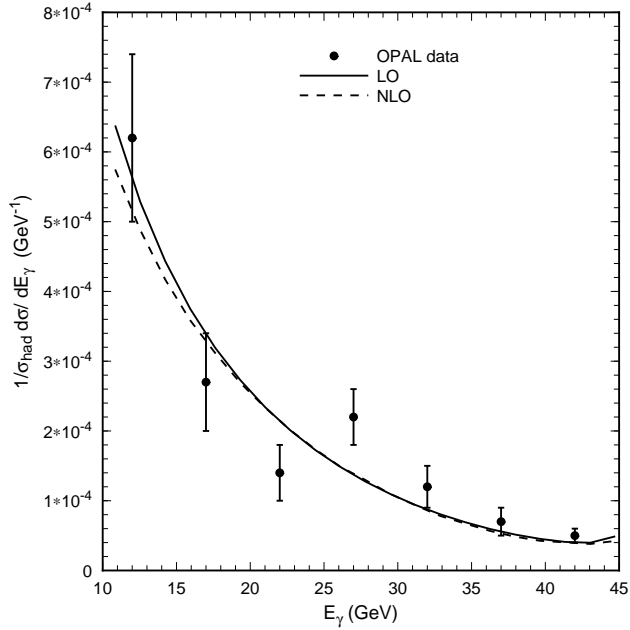,width=9cm}
\caption{The inclusive photon energy distribution 
normalized to the hadronic cross section as measured by the OPAL Collaboration
compared with the ${\cal O}(\alpha)$ and ${\cal O}(\alpha \alpha_s)$
order calculations including the appropriate 
quark-to-photon fragmentation functions 
determined using the ALEPH photon + 1~jet data.}
\label{fig:inclusive}
\end{center}
\end{figure}

As a final illustration of the generality of our results, 
we turn to the inclusive photon distribution recently 
measured by the OPAL Collaboration
\cite{OPALinc}.
Here, jets are not defined and, rather than 
the distribution of electromagnetic energy 
within the cluster, the energy distribution of the 
photon is considered.   OPAL are able to identify photons with energies 
with as little as 10~GeV, and have compared their results with the model
estimates of~\cite{owens,grv}.  They find reasonable agreement 
with factorization scales $\mu_F \sim M_Z$, although the data 
is too poor to discriminate between the models.  
Fig.~\ref{fig:inclusive} shows our (scale independent) predictions 
for the inclusive photon energy distribution at both leading and 
next-to-leading order.
We see good agreement with the data, even though the phase space 
relevant for the OPAL data 
far exceeds that used to determine the fragmentation functions from the ALEPH
photon + 1 jet data.

In summary, we have presented the results of a complete calculation~\cite{big} 
of the `photon' +~1 jet rate at ${\cal O}(\alpha\alpha_s)$. Although 
only next-to-leading order in perturbation theory, this calculation 
contains several ingredients appropriate to the calculation of 
jet observables at next-to-next-to-leading order. In particular, we 
needed to generalize the phase space slicing method of~\cite{kramer,gg}
to more than one theoretically unresolved particle and
have analytically calculated the following double unresolved 
configurations~\cite{big}:
the triple collinear factor, the soft/collinear factor and the 
double single collinear factor. The `photon' +~1 jet rate was then evaluated 
for a democratic clustering algorithm with a Monte Carlo program using 
the hybrid subtraction method of~\cite{eec}.
The results of our calculation, when compared to the data~\cite{aleph}
on the `photon' +~1 jet rate obtained by ALEPH, 
enabled a first determination
of the process independent quark-to-photon fragmentation function
at ${\cal O}(\alpha\alpha_s)$ in a fixed order approach. 
As a first application, we have used this 
function to calculate the `isolated' photon +~1 jet rate in a democratic 
clustering approach at next-to-leading order. The inclusion of the QCD
corrections improves the agreement between theoretical
prediction and experimental data.
Moreover, it was shown that these corrections are moderate,
demonstrating 
the perturbative stability of this particular isolated photon definition.
Finally, we have computed the inclusive photon energy distribution and 
found good agreement with the recent OPAL data.

\section*{Acknowledgements}
One of us (A.G.) would like to thank the DESY Theory Group 
for their kind hospitality during the later stages of this work.
A.G. also acknowledges the 
financial support of the University of Durham 
through a Research Studentship (Department of Physics).

\end{document}